\documentclass [aps,prl,amsmath,amssymb,twocolumn, letterpaper]{revtex4}
\usepackage{amsmath}
\usepackage{graphicx}

\begin{document}

\title{Phase-incoherent superconducting pairs in the normal state of Ba(Fe$_{1-x}$Co$_x$)$_2$As$_2$ }

\author{G. Sheet,$^1$ M. Mehta,$^1$ D. A. Dikin,$^1$ S. Lee,$^2$ C.W. Bark,$^2$ J. Jiang,$^3$\\
J. D. Weiss,$^3$ E. E. Hellstrom,$^3$ M.S. Rzchowski,$^4$ C.B. Eom$^2$ and V. Chandrasekhar$^1$ }

\affiliation{$^1$Department of Physics and Astronomy, Northwestern University, Evanston, IL 60208, USA,\\ 
 $^2$Department of Materials Science and Engineering University of Wisconsin-Madison, Madison, WI 53706, USA,\\ $^3$Applied Superconductivity Center, National High Magnetic Field Laboratory, Florida State University, Tallahassee, FL 32310, USA, \\$^4$Physics Department
University of Wisconsin-Madison,Madison, WI 53706  }

\begin{abstract}

 The normal state properties of the recently discovered ferropnictide superconductors might hold the key to understanding their exotic superconductivity. Using point-contact spectroscopy we show that Andreev reflection between an epitaxial thin film of Ba(Fe$_{0.92}$Co$_{0.08}$)$_2$As$_2$ and a silver tip can be seen in the normal state of the film up to temperature $T\sim1.3T_c$, where $T_c$ is the critical temperature of the superconductor. Andreev reflection far above $T_c$ can be understood only when superconducting pairs arising from strong fluctuation of the phase of the complex superconducting order parameter exist in the normal state. Our results provide spectroscopic evidence of phase-incoherent superconducting pairs in the normal state of the ferropnictide superconductors. 

\end{abstract}

\maketitle

Ever since the ferropnictide superconductors were discovered as a new class of high temperature superconductors \cite{kamihara}, there has been speculation about the similarity of their phase diagram with that of the high $T_c$ cuprates and the pairing symmetry of the superconducting state.  The most widely accepted scenario for the origin of superconductivity in the ferropnictide superconductors is described within a multiband picture where both electron and hole pockets in the Fermi surface contribute, giving rise to an extended $s$-wave symmetry, the so-called $s$+$-s$- symmetry \cite{Mazin}.  In this respect the ferropnictides differ from the high $T_c$ cuprates, where an anisotropic superconducting gap originates from a single band in the Fermi surface. In terms of the phase diagram, the ferropnictides, particularly the BaFe$_2$As$_2$ family \cite{Chu}, show strong similarities with the high $T_c$ cuprate superconductors.  Most importantly, the parent material is antiferromagnetic and spin fluctuations might play a crucial role in pairing \cite{Chu, Hess}. 

A number of recent experiments have reported observing a pseudogap \cite{Mertelj,Ahilan,Sato,Pan} in the ferropnictide superconductors. The pseudogap in the superconductors with low superconducting carrier density such as in the underdoped cuprates is thought to arise from fluctuations in the phase of the complex superconducting order parameter  that give rise to phase-incoherent quasiparticle pairs well above $T_c$ \cite{Emery,Davis,Deutscher}.  However, since the undoped ferropnictides are semimetals (as opposed to the cuprates whose parent compounds are Mott insulators), the superconducting carrier density in the ferropnictides is large and comparable to that of the conventional superconductors \cite{Luan}. Thus, ideally phase fluctuations in the ferropnictides should not be significant \cite{Emery}, raising questions about the origin of the pseudogap in the ferropnictides.  Here we report on our experiments probing the nature of the normal state of the ferropnictide superconductor Ba(Fe$_{1-x}$Co$_x$)$_2$As$_2$, using point contact Andreev reflection (PCAR) spectroscopy \cite{Choi,Deutscher}. From PCAR measurements on high quality epitaxial thin films of Ba(Fe$_{1-x}$Co$_x$)$_2$As$_2$, we show that a signature of superconducting correlations due to phase-incoherent pairing can be seen at temperatures more than 30\% greater than the superconducting transition temperature $T_c$ of the film, far above the temperatures at which conventional superconducting fluctuations are expected to vanish \cite{Tinkham}.  

\begin{figure}
\includegraphics[width=9 cm]{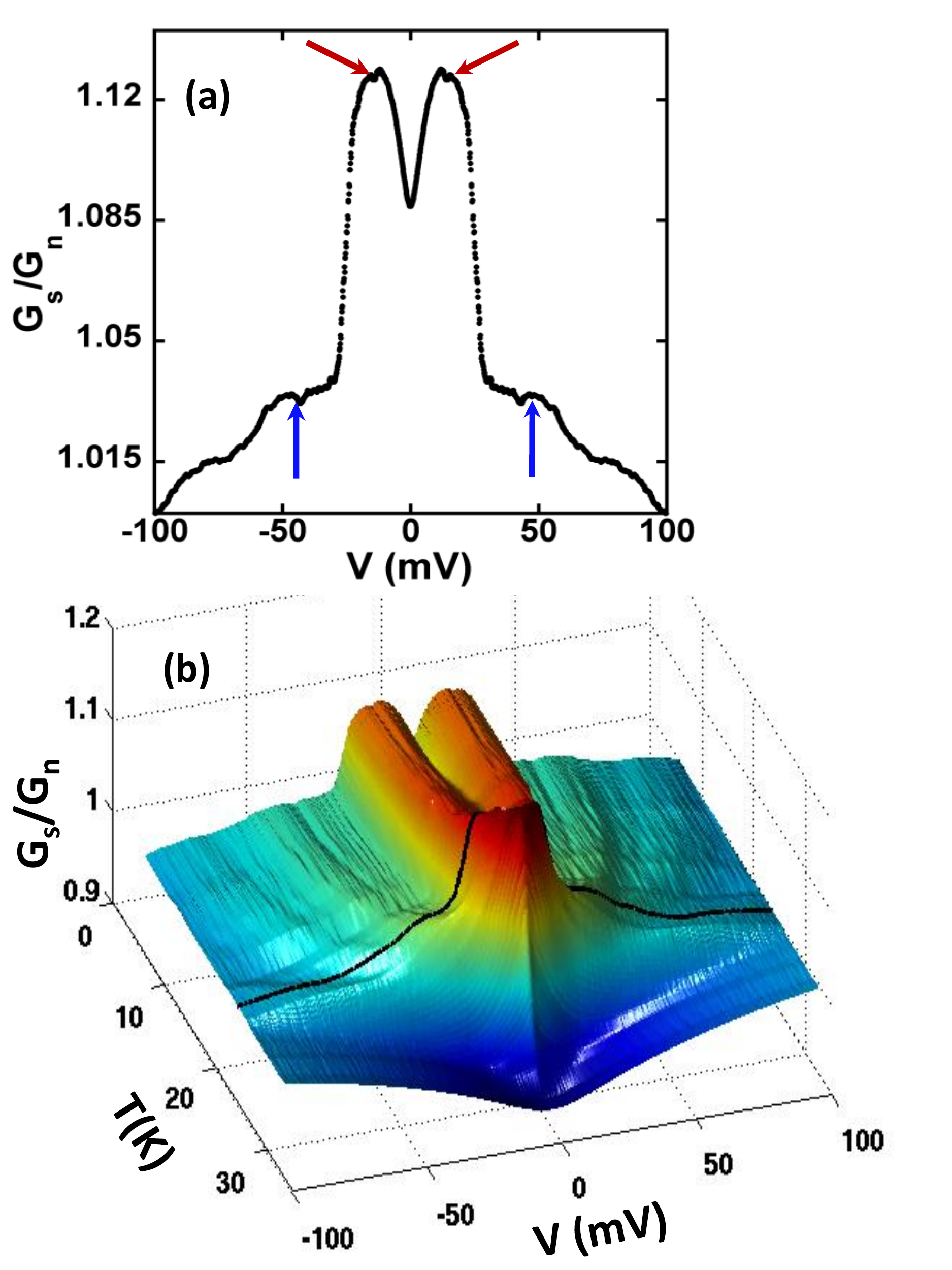}
\caption{Point contact data on a Ba(Fe$_{1-x}$Co$_x$)$_2$As$_2$ epitaxial film with $x \sim 0.08$. (a) The symmetrized conductance $G_s$ normalized to its value $G_N$ at 100 mV at $T=10.4$ K. (b)  Point contact spectra as a function of temperature $T$.  The black line in (b) corresponds to the data at $T=23.7$ K, the mid-point of the resistive transition. 
\label{fig1} }
\end{figure}

\begin{figure}[htb]
\includegraphics[width=8.5 cm]{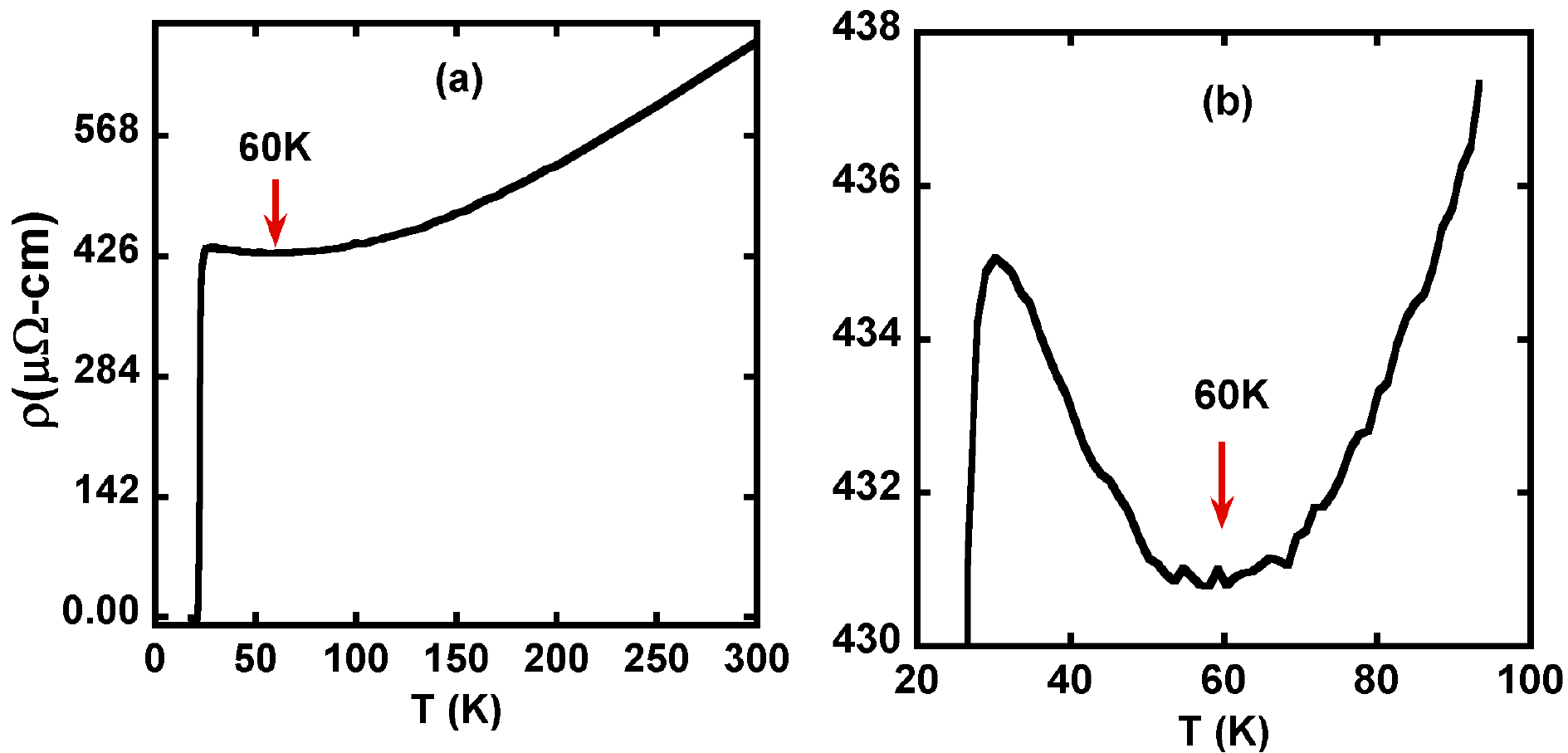}
\caption{(a) Four probe resistivity of the epitaxial film as a function of temperature. (b) The temperature dependence of resistivity close to the transition.  
\label{fig2} }
\end{figure}

The samples used for this study are very high quality single crystal films \cite{Eom2}. All the spectra for different point-contacts show an asymmetry with respect to voltage across the point-contact. For the analysis of these spectra, we extract the symmetric and the antisymmetric parts of the measured differential resistance ($dV/dI$) with respect to the voltage as follows
\begin{equation*}
(dV/dI)_{s,a} =\frac{(dV/dI) [V] \pm (dV/dI) [-V]}{2}
\label{eqn1}
\end{equation*}
where $V$ is the voltage across the point contact.
Here we concentrate on the analysis of the symmetric conductance $G_s = 1/(dV/dI)_s$, a representative trace of which is shown in Fig. \ref{fig1}a. The temperature dependence of the symmetric part of the point contact conductance is presented in a three-dimensional plot in Fig. \ref{fig1}b.   Data for each temperature are normalized to the conductance $G_N$ at a voltage of 100 mV. At lower temperatures, two major peaks in $G_s$ centered at $\sim\pm$12 mV are observed, with a dip in conductance at zero bias (Fig. \ref{fig1}a).  In addition, a number of smaller amplitude features are seen. As the temperature is increased, the two major peaks move to lower values of the voltage bias; at temperatures above $\sim$28 K, the most prominent feature is a peak in $G_s$ at zero bias that eventually disappears as the temperature is increased further, leaving a broad background that is temperature independent above $\sim$32 K.  The most striking feature of the point-contact spectra shown in Fig. \ref{fig1}b is that the voltage dependent point-contact conductance shows strong modulation \textit{above} $T_c$.  $T_c$ for this film is determined from the midpoint of the resistive transition to be 23.7 K (see Fig. \ref{fig2}). The point-contact spectrum for this temperature (23.7 K) is shown as the solid black line in Fig. \ref{fig1}b.  

In order to put this observation in context we will briefly review the analysis of typical point contact spectra on superconductors. Point contact spectra of conventional $s$-wave superconductors, $d$-wave high $T_c$ cuprate superconductors and even two-band superconductors such as MgB$_2$ have been analyzed successfully using adaptations of the model of Blonder, Tinkham and Klapwijk (BTK) \cite{BTK}. For the simplest case of a single-band $s$-wave superconductor, the point contact spectra depend on the superconducting gap $\Delta$, the transparency of the contact between the normal tip (N) and the superconductor (S), which is usually characterized by the BTK parameter $Z$ ($Z$=0 for perfectly transparent contacts, and $Z \rightarrow \infty$ in the tunneling regime), and the inelastic lifetime $\Gamma$, which is usually put in by hand as the complex component of the quasiparticle energy $E$, $E\rightarrow E + i\Gamma$.  For $Z=0$ and $T<<T_c$, the bias dependent conductance increases by a factor of 2 in going from $|V|>\Delta/e$ to $|V|<\Delta/e$, corresponding to the factor of two enhancement of conductance that arises from Andreev reflection at the N-S interface.  As $Z$ increases, the probability of Andreev reflection decreases, resulting in a progressive drop in the zero bias conductance, until $G_s$ vanishes at zero bias as $Z\rightarrow \infty$.  For finite $Z$, this results in two peaks in the conductance at voltages comparable to $\pm \Delta/e$.  This is qualitatively what is observed in the data presented in Fig. \ref{fig1}. We have ignored all spectra showing high-bias conductance dips \cite{Sheet} and spectra without the zero-bias conductance dips (which is the hallmark of a ballistic point-contact), and analyzed only those in the ballistic limit. 

A visual inspection of the 10.4 K point contact spectrum (Fig. \ref{fig1}a) reveals that the low-bias peaks symmetric about $V$ = 0 are significantly broader than what is expected for a single band superconductor. This broadening might be due to existence of two gaps close to each other that are not well resolved. A small kink (indicated by the red arrows in Fig. \ref{fig1}a) further indicates that two closely located gaps might indeed exist. This is similar to earlier observations in the ferropnictide superconductors \cite{Yin,Ding,Gonnelli,Chen}. The presence of the two gaps is even more pronounced in other point contact spectra that we have measured. Based on this, we have attempted to fit the spectra to a two-band BTK model.  However,  the quality of the fits is not very good, and the values of $\Gamma$ are very large, comparable to $\Delta$, which is unphysical.

The temperature evolution of the spectra is unusual. In point contact measurements on a conventional $s$-wave single-band superconductor with finite $Z$, the conductance peaks at finite bias move to lower values of voltage with increasing temperature, and the low bias conductance decreases, with all spectral features vanishing at $T_c$.  In our data, on the other hand, the low-bias conductance increases with increasing temperature upto 27 K (which is greater than $T_c$), after which it starts dropping.  The overall shape of the spectrum also starts changing: the double peak structure vanishes at 25 K, leaving a sharp cusp at zero bias at higher temperatures. Even at 30 K, a small zero bias cusp is observed signifying that there is Andreev reflection at a temperature that is significantly higher than the bulk $T_c$. 

The observation of Andreev reflection far above $T_c$ has been tied to the presence of phase-incoherent superconducting pairs \cite{Choi,Deutscher,Chang} that is thought to give rise to the pseudogap in the cuprates. Within the model of phase-incoherent pairs originating from the classical fluctuation of the phase of the superconducting order parameter of the superconductors with low super fluid density, pairing of the quasiparticles is expected to initiate at a temperature much higher than $T_c$, but global phase coherence is not achieved until the material is cooled down to $T_c$.  This is in contrast to the conventional superconductors with high superfluid density where the pair formation and phase-coherence coincide at $T_c$. A signature of phase-incoherent superconductivity in the normal state of the cuprates was observed experimentally \cite{Davis}, but Andreev reflection involving phase-incoherent pairs was never observed in the cuprates \cite{Deutscher,Chang}.  Although it is believed that the superfluid density in the ferropnictides is comparable to that of the conventional superconductors \cite{Luan}, our observation of the phase-incoherent superconducting pairs in the normal state of the ferropnictide superconductor  Ba(Fe$_{1-x}$Co$_x$)$_2$As$_2$ indicates that stiffness of the phase of the complex superconducting order parameter in the ferropnictides is not as high as expected.

It should be noted that in principle, conventional thermal fluctuations of the order parameter may give rise to Andreev reflection and consequent zero-bias conductance enhancement above $T_c$ even for a conventional BCS superconductor with disorder \cite{Tinkham}. However, such fluctuations have a very small contribution to the conductivity, and Andreev reflection due to such fluctuations was never observed for conventional BCS superconductors. A small contribution of Andreev reflection associated with such fluctuations has been reported for overdoped cuprates \cite{Chang}, but this contribution survives only up to $T = 1.01T_c$. Therefore, conventional thermal fluctuations are not expected to contribute to the large zero-bias enhancement that we observe up to $T = 1.3T_c$. Other possible orderings in the normal state such as a charge density wave (CDW) or a spin-density wave (SDW) would result in completely different behavior, i.e., the Fermi surface nesting associated with such ordering would cause a reduction in conductance in the normal state \cite{Choi}.   On the other hand, the fact that the conductance enhancement in this case arises from Andreev reflection involving phase-incoherent pairs is strongly supported by the following facts: (1) the low-temperature Andreev reflection spectra evolves smoothly with temperature up to 30 K without any noticeable change at $T_c$; (2) a spectral feature is seen at high bias ($\sim$ 50 mV) that evolves with temperature in a fashion similar to the low-bias peaks, and merges smoothly with the low-bias peaks at higher temperatures before vanishing at 30 K;  and (3) the numerical simulations involving a high-bias gap that participates in Andreev reflection above $T_c$ reproduce all the experimental observations, as we shall show below. We have observed similar behavior on several different points on the sample whose data are shown in Fig. \ref{fig1}, and also on another sample with a slightly smaller $T_c$ = 18.5 K where the lower $T_c$ arises from a higher level of disorder. 

The temperature dependence of resistivity for the sample is shown in Fig. \ref{fig2}.  In the high temperature regime, the resistivity decreases almost linearly with decreasing temperature and deviates from linear behavior at a temperature of $T\sim$180 K. It shows a superconducting transition at 23.7 K ($T_c$ is defined as the midpoint of the resistive transition) with a transition width of 1.3 K.  
This data does not show any signature of antiferromagnetic ordering \cite{Chu}. This gives further support for the fact that the enhancement in conductance that we observe above $T_c$  cannot be attributed to such magnetic ordering in the normal state. A close inspection of the resitivity data above the superconducting transition reveals that the resistivity starts increasing with decreasing temperature  giving rise to a broad minimum in resistivity at $T\sim$ 60 K (Fig.\ref{fig2}b). This upturn was attributed to a structural phase transition in the slightly underdoped ferropnictides \cite{Chu} which is not the case here, as discussed before. However, this behavior of the normal state resistivity is strikingly similar to what has been observed in high $T_c$ cuprate superconductors \cite{Daou}.  In the context of the cuprates, this upturn was associated with a decreased quasiparticle density of states in the Fermi surface leading to a pseudogap. Formation of incoherent pairs in the normal state of the ferropnictides could also cause such a reduction in the quasiparticle density of states at the Fermi level.

\begin{figure}[htb]
\includegraphics[width=8.5 cm]{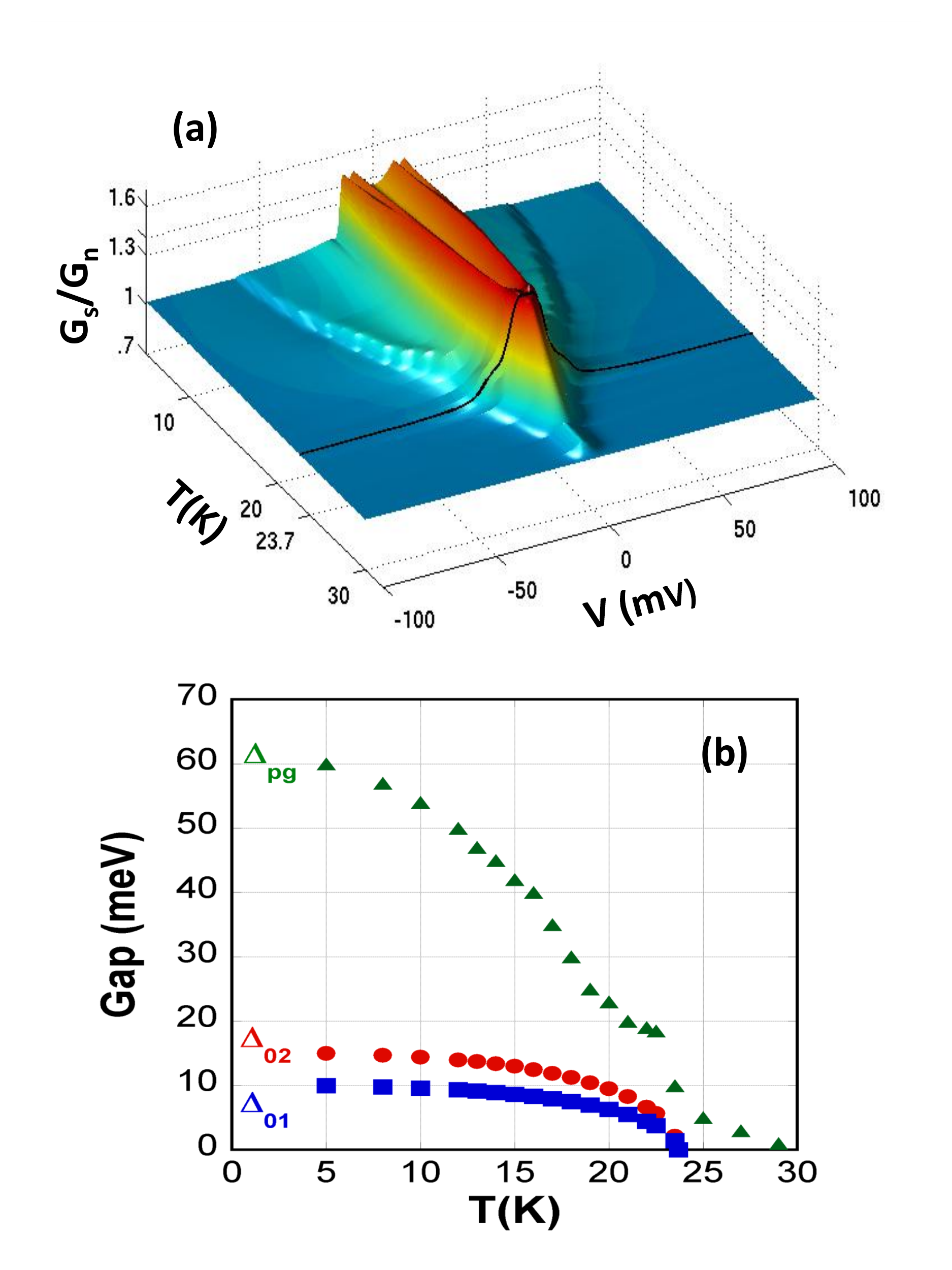}
\caption{(a) Simulation of the point-contact conductance of a superconductor with three gaps including a normal state gap using a modified BTK model as described in the text. (b) The blue and the red dots indicate the assumed BCS-like gaps and the green dots indicate the high bias gap used for the simulation. The black line shows the calculated curve at 23.3 K, close to the assumed $T_c$.
\label{fig3}}
\end{figure}

In order to illustrate how a normal state gap that participates in Andreev reflection might give rise to the observed temperature dependence, we have simulated spectra using a modified BTK model with two BCS-like \cite{Tinkham} gaps ($\Delta_{01}$ and $\Delta_{02}$) at low bias that vanish at $T_c$, and another ($\Delta_{pg}$) at high bias and studied the temperature dependence of the entire spectrum numerically. The contributions of the three gaps $\Delta_{01}$, $\Delta_{02}$ and $\Delta_{pg}$ to the point contact spectra was taken in the ratio of 3:6:1, as one might expect the phase-incoherent quasiparticle pairs to contribute a much smaller amount to the Andreev conductance than the coherent pairs in the superconducting state. $\Delta_{pg}$ was varied to obtain spectra that looked qualitatively like the experimental data (Fig. \ref{fig3}b). Figure \ref{fig3}a shows the resulting spectrum.  It should be noted that the experimental data show a background that is temperature independent above $\sim$32 K, and this background is not included in the simulation. $\Delta_{pg}$ has a much stronger temperature dependence than the BCS-like gaps below $T_c$, shows a sharp drop near $T_c$  and is non-zero up to a temperature of 30 K, significantly higher than the temperature where $\Delta_{01}$ and $\Delta_{02}$ vanish.  While the details of the curves depend on the assumptions and the values of the parameters, the qualitative behavior is the same over a wide range of parameter values.   In particular, in the low temperature regime, the low-bias conductance increases with increasing temperature, then decreases above a certain temperature (26 K in this case) and finally develops a cusp at $V=0$ as the temperature is increased further. This behavior is strikingly similar to our experimental observation.  Consequently, the numerical simulations that assume the existence of a normal state gap involving phase-incoherent superconducting pairs and two low-energy superconducting gaps qualitatively reproduce our experimental results. 

In conclusion, we have provided evidence for Andreev reflection in the metallic point-contacts on epitaxial thin films of the ferropnictide superconductor Ba(Fe$_{0.92}$Co$_{0.08}$)$_2$As$_2$. This clearly indicates the formation of phase-incoherent quasiparticle pairs at a temperature well above $T_c$. Based on these results, we believe that the role of fluctuations in superconducting ferropnictides needs to be explored both theoretically and experimentally.

We acknowledge fruitful discussions with J. Sauls, M. Norman, A. Chubukov and A. Vorontsov. U.S. Department of Energy supported the work at Northwestern University through Grant No. DE-FG02-06ER46346 and at the University of Wisconsin through Grant no.  DE-FG02-06ER46327, while the work at the NHMFL was supported under NSF Cooperative Agreement DMR-0084173, by the State of Florida, and by AFOSR under grant FA9550-06-1-0474.

\end{document}